\documentclass[aps,pra,epsfig,showpacs,superscriptaddress,footnote,twocolumn]{revtex4}
\makeatletter
\usepackage{calc}
\usepackage{times}
\usepackage{amsmath, amssymb}
\usepackage{multirow}
\def\be{\begin{equation}}
\def\ee{\end{equation}}
\def\bea{\begin{eqnarray}}
\def\eea{\end{eqnarray}}
\def\bma{\begin{mathletters}}
\def\ema{\end{mathletters}}
\def\bi{\begin{itemize}}
\def\ei{\end{itemize}}

\def\C{\hbox{$\mit I$\kern-.7em$\mit C$}}

\newcommand{\singlespacing}{\let\CS=\@currsize\renewcommand{\baselinestretch}
{1.0}\tiny\CS}
\newcommand{\doublespacing}{\let\CS=\@currsize\renewcommand{\baselinestretch}
{1.5}\tiny\CS}

\begin{document}

\title{Local randomness in Hardy's correlations: Implications from information causality principle. }

\author{MD. Rajjak Gazi}
\email{rajjakgazimath@gmail.com} \affiliation{Physics and applied
mathematics unit, Indian statistical unit, 203 B.T. Road, Kolkata-700108,
India}

\author{Ashutosh Rai}
\email{arai@bose.res.in} \affiliation{S.N.Bose National Center for
Basic Sciences,Block JD, Sector III, Salt Lake, Kolkata-700098,
India}

\author{Samir Kunkri}
\email{skunkri@yahoo.com} \affiliation{Mahadevananda
Mahavidyalaya, Monirampore, Barrackpore, North 24 Parganas,
700120, India }

\author{Ramij Rahaman}
\email{ramij.rahaman@ii.uib.no} \affiliation{Selmer Center,
Department of Informatics, University of Bergen, Bergen, P.O. Box-7803, N-5020, Norway}

\begin{abstract}
Study of nonlocal correlations in term of Hardy's argument has been quite popular in quantum mechanics.
Recently Hardy's argument of non-locality has been studied in the context of generalized non-signaling
theory as well as theory respecting information causality. Information causality condition significantly
reduces the success probability for Hardy's argument when compared to the result based on non-signaling condition.
Here motivated by the fact that maximally entangled state in quantum mechanics
does not exhibit Hardy's non-local correlation,
we do a qualitative study of the
property of local randomness of measured observable on each side reproducing Hardy's non-locality correlation,
in the context of information causality condition.
On applying the necessary condition for respecting the principle of
information causality, we find that there are severe restrictions on the
local randomness of measured observable in contrast to results obtained from  no-signaling condition.
Still, there are some restrictions imposed by quantum mechanics that are not obtained from information causality condition.
\end{abstract}

\maketitle
\section{Introduction}
 Violation of the Bell-type inequalities \cite{bell} by quantum mechanics show that
nature is nonlocal. Nevertheless quantum correlations respect
causality principle \cite{ghirardi}. However, there are also other
non-signaling post quantum correlations \cite{PR} which cannot be
distinguished from quantum correlation by subjecting them to the
causality principle. Though post quantum correlations are not
observed in experiments, but still we do'nt understand what
underlying physical principle(s) completely distinguishes quantum
correlations from nonphysical post quantum correlations.

Recent studies has shown that quantum features like violation of
Bell type inequalities \cite{PR}, intrinsic randomness, no-cloning
\cite{a,b}, information-disturbance tradeoff \cite{c}, secure
cryptography \cite{d,e,f}, teleportation \cite{g}, entanglement
swapping \cite{h} are also enjoyed by other post quantum
no-signaling theories. On the other hand for no-signalling
correlations some implausible features has also been noticed like:
some no-signalling correlations would make certain distributed
computational tasks trivial \cite{i,j,k,l} and would have very
limited dynamics \cite{m}. So the study of the nonlocal
correlations in the general no-signaling framework
\cite{a,b,c,d,e,f,g,h,i,j,k,l,m,n} leads us towards a deeper
understanding of quantum correlations.

Very recently, non-violation of information causality (IC)
\cite{IC} has been identified as one of the foundational principle
of nature, it is compatible with experimentally observed quantum
and classical correlations but rules out an unobserved class of
nonlocal correlation as nonphysical. The principle states that
communication of $m$ classical bits causes information gain of at
most $m$ bits, this is a generalization of the no-signalling
principle, the case $m=0$ corresponds to no-signalling. Applying
IC principle to non-local correlations, we get the Tsirelson's
bound \cite{tsirelson} and all correlations that goes beyond
Tsirelson's bound violate the principle of information causality
\cite{IC}. In \cite{allcock} it was shown that though some part of
quantum boundary can be derived from a necessary condition (given
in \cite{IC}) for violating IC, this condition is not sufficient
for distinguishing quantum correlations from all post-quantum
correlations which are below the Tsirelson's bound. So it remains
interesting to see if the full power of IC (some other conditions
derived from IC) can eliminate remaining post-quantum correlations
below the Tsirelson's bound. Along with the research in the
direction of completely distinguishing the quantum correlations
from rest of the nonlocal correlations, it would also be
interesting to apply the known IC condition(s) for
qualitative/quantitative study of certain specific features of
nonlocal correlations. For instance, it was known that maximum success probability of Hardy's nonlocality
argument \cite{hardy92,hardy93} under the no-signaling restriction is $0.5$ \cite{hardyns} and
within quantum mechanics the maximum takes the value $0.09$ \cite{hardyqm}, then by applying
the IC principle, in \cite{ashu} it was shown that the upper bound on success probability reduces to $0.20717$.

In the present article we apply IC condition in order to study
the property of local randomness for a bipartite probability
distribution which exhibits Hardy's non-locality
\cite{hardy92,hardy93}. Our motivation for this study came from
the fact that Hardy's non-locality argument in quantum mechanics
does not work for maximally entangled state \cite{hardy93,cabello}
and at the same time for a maximally entangled state, local
density matrix being completely random, both the results for a
qubit are equally probable. Keeping this in mind, we asked a more
general question like: for two two-level systems, how many
observable and in which way, out of four entering in the Hardy's
non-locality argument, can be locally random. We want to study
this question in the context of probability distribution which
respects an IC condition as well as in the context of quantum
mechanics. We see that the applied IC condition itself imposes
powerful restriction but still it does not reproduce all the
restrictions imposed by quantum mechanics. In this context, it is
to be mentioned that no signalling condition does not impose any
such restriction. Interestingly we observed that the applied
necessary condition for respecting IC allows at most two
observable, one on each side, chosen in a restricted way to be
completely random, and quantum mechanics allows only one of them
to be completely random.

This article is organized as follows. In
Sec. II we discuss the general structure of the set of a bipartite
two input-two output nonsignaling correlations. In Sec. III we
restrict the the type of correlations in Sec. II by Hardy's
nonlocality conditions. In Sec. IV we study the property of local
randomness in Hardy's correlation, in Sec. IV A we make this study
for no-signaling correlations, in Sec. IV B we study it for
correlations respecting an IC condition, in Sec. IV C we work it
for quantum correlations. We give our conclusions in Sec. V.

\section{Bipartite nonsignaling correlations}
Let us consider a bipartite black box shared between two parties:
Alice and Bob. Alice and Bob input variables $x$ and $y$ at their
end of the box, respectively, and receive outputs $a$ and $b$. For
a fixed input variables there can be different outcomes with
certain probabilities. The behavior of a these correlation boxes
is fully described by a set of joint probabilities $P(ab|xy)$. In
this article, we will focus on the case of binary inputs and
outputs $(a,b,x,y \in \{0,1\})$. Then we have a set of $16$ joint
probabilities defining a bipartite binary input - binary output
correlation box. These types of correlations can be
represented by a $4\times 4$ correlation matrix:\\

\begin{center}
$\left(
                 \begin{array}{cccc}
                  P(00|00)&P(01|00)&P(10|00)&P(11|00)\\
P(00|01)&P(01|01)&P(10|01)&P(11|01)\\
P(00|10)&P(01|10)&P(10|10)&P(11|10)\\
P(00|11)&P(01|11)&P(10|11)&P(11|11)\\
                 \end{array}
               \right)$
\end{center}

We note that since $P(ab|xy)$ are probabilities, they satisfy
positivity, $P(ab|xy)\geq 0 \; \forall \; a,b,x,y,$ and
normalization $\sum _{a,b}P(ab|xy)= 1 \; \forall \; x,y.$ Since we
are to study nonsignaling boxes; i.e., we require that Alice
cannot signal to Bob by her choice $x$ and vice versa, the
marginal probabilities $P_{a|x}$ and $P_{b|y}$ must be independent
of $y$ and $x$, respectively. The full set of nonsignaling boxes
forms an eight-dimentional polytope \cite{n} which has $24$
vertices: eight extremal nonlocal boxes and $16$ local
deterministic boxes. The extremal nonlocal correlations have the
form
\begin{equation}
\label{nonlocal} P_{NL}^{\alpha \beta \gamma } =
\left\{\begin{array}{ccccc}
                   \frac{1}{2} & {\rm if} & a \oplus b &=& XY \oplus {\alpha}X \oplus {\beta}Y \oplus \gamma,\\
                   0 & {\rm otherwise}, & & &
                   \end{array}
             \right.
\end{equation}
where $\alpha, \beta, \gamma \in \{ 0, 1\}$ and $\oplus$ denotes
addition modulo $2.$. Similarly, the local deterministic boxes are
described by
\begin{equation}
\label{local} P_{L}^{\alpha \beta \gamma \delta} =
\left\{\begin{array}{ccccc}
                   1 & {\rm if} & a &=& {\alpha}X \oplus {\beta},\\
                     &           & b &=& {\gamma}Y \oplus {\delta};\\
                   0 & {\rm otherwise}, & & &
                   \end{array}
             \right.
\end{equation}
where $\alpha, \beta, \gamma, \delta \in \{ 0, 1\}$ and $\oplus$
denotes addition modulo $2.$\\
Thus we can see that any bipartite two input- two output
nonsignaling correlation box can be expressed as
a convex combination of the above $24$ local/nonlocal vertices.\\

\section{Hardy's correlations under no-signaling condition}
A bipartite two input - two output Hardy's correlation puts simple
restrictions on a certain choice of $4$ out of $16$ joint
probabilities in the correlation matrix. One such choice is
$P(11|11)>0$, $P(11|01)=0$, $P(11|10)=0$, $P(00|00)=0$ and it is
easy to argue that these correlations are nonlocal. To show this,
let us suppose that these correlations are local i.e. they can be
simulated by noncommunicating observers with only shared
randomness as a resource. Now consider the subset of those random
variables $\lambda$ shared between the two observers such that for
$\lambda$s belonging to this subset input $x=1,y=1$ give output
$a=1,b=1$ (this subset is nonempty since $P(11|11)>0$), now
conditions $P(11|01)=0$ and $P(11|10)=0$ tell that within this
subset input $x=0,y=0$ would give output $a=0,b=0$, this would
imply that $P(00|00)>0$), but it contradicts the condition
$P(00|00)=0$. Hence these correlations are nonlocal. If we further
restrict these correlations by no-signaling condition we get
Hardy's nonsignaling boxes. It is easy to check that these boxes
can be written as a convex combination of $5$ of the sixteen local
vertices $P_{L}^{0001}$, $P_{L}^{0011}$, $P_{L}^{0100}$,
$P_{L}^{1100}$, $P_{L}^{1111}$ and $1$ of the eight nonlocal
vertex $P_{NL}^{001}$. Then,
\begin{eqnarray}
\label{hardy} P_{ab|XY}^{{\cal H}} &=& c_1 P_{L}^{0001} + c_2
P_{L}^{0011} + c_3 P_{L}^{0100} \nonumber \\
&&+ c_4 P_{L}^{1100}+ c_5 P_{L}^{1111} + c_6 P_{NL}^{001}
\end{eqnarray}
where $\sum_{j = 1}^{6} {c}_i = 1.$ From here the correlation matrix for these Hardy's nonsignaling
boxes can be written as\\
\begin{center}
$\left(
\begin{array}{cccc}
0 & c_{1}+c_{2}+\frac{c_{6}}{2} & c_{3}+c_{4}+\frac{c_{6}}{2} & c_{5}\\
c_{2}&c_{1}+\frac{c_{6}}{2}&c_{3}+c_{4}+c_{5}+\frac{c_{6}}{2}& 0\\
c_{4}&c_{1}+c_{2}+c_{5}+\frac{c_{6}}{2}&c_{3}+\frac{c_{6}}{2}&0\\
c_{2}+c_{4}+c_{5}+\frac{c_{6}}{2}&c_{1}&c_{3}&\frac{c_{6}}{2}\\
\end{array}
               \right)$
\end{center}

\section{Property of local randomness in Hardy's correlations}
For a most general bipartite correlation an input $x$ on Alice's
side is locally random if the marginal probabilities of all
possible outcomes on Alice's side for this input, are equal and
similarly for Bob. In the case of two input- two output bipartite
correlations: an input $x$ on Alice's side is locally random if,
$P(0|x)=P(1|x)=\frac{1}{2}$, in terms of joint probabilities this
would mean that for any choice of Bob's input $y$,
$P(00|xy)+P(01|xy)=P(10|xy)+P(11|xy)= \frac{1}{2}$. Similarly an
input $y$ on Bob's side is locally random if,
$P(0|y)=P(1|y)=\frac{1}{2}$, in terms of joint probabilities this
can be expressed as, for any choice of Alice's input $x$,
$P(00|xy)+P(10|xy)=P(01|xy)+P(11|xy)= \frac{1}{2}$. Let us denote
the $0$ and $1$ inputs on Alice's (Bob's) side as $0_{A}$($0_{B}$)
and $1_{A}$($1_{B}$) respectively. We would now like to see that,
what choices of inputs from the set $\{0_{A}, 1_{A}, 0_{B},
1_{B}\}$ can be locally random for a given class of Hardy's
correlations.

\begin{table}[h!]

\begin{tabular}{|c|c|}
  \hline
  Input& Conditions for local randomness \\
  \hline \hline
  $0_{A}$ & $c_1+c_2+\frac{c_6}{2}=\frac{1}{2}$ \\& $c_3+c_4+c_5+\frac{c_6}{2}=\frac{1}{2}$ \\
  \hline
  $1_{A}$ &  $c_1+c_2+c_4+c_5+\frac{c_6}{2}=\frac{1}{2}$ \\ & $c_3+\frac{c_6}{2}=\frac{1}{2}$\\
  \hline
  $0_{B}$ & $c_3+c_4+\frac{c_6}{2}=\frac{1}{2}$\\ & $c_1+c_2+c_5+\frac{c_6}{2}=\frac{1}{2}$ \\
  \hline
  $1_{B}$ & $c_2+c_3+c_2+c_4+c_5+\frac{c_6}{2}=\frac{1}{2}$ \\ & $ c_1+\frac{c_6}{2}=\frac{1}{2}$\\
  \hline
\end{tabular}
\caption{For the no-signaling bipartite Hardy's correlation with
two dichotomic observable on either side, here each row give the
conditions which coefficients $c_{i}$s must satisfy for the
corresponding input to be locally random.}
\end{table}

\subsection{Hardy's correlations respecting no-signaling}
In the case of Hardy's correlations which respects no-signalling,
condition of local randomness for each of the possible inputs, are given in the TABLE I.
Now let us see that for the Hardy's correlations respecting
no-signalling, what choices of inputs can be locally random. We
give the results for every case, in the TABLE II. We can read from
here that although in order to show the property of local
randomness Hardy's correlations becomes much restricted, yet we
get solutions for each case. If we get solutions for the case 1, it is obvious that there are solutions in all the remaining cases 2-15 ,
nevertheless we write the complete table giving the form of solutions in each case
for the later reference.
\begin{widetext}{\tiny
\begin{table}[h!]
\begin{center}
\begin{tabular}{|c|c|c|c|c|c|c|c|}
  \hline
  Cases & Locally random inputs &$C_1$&$C_2$&$C_3$&$C_4$&$C_5$&$C_6$\\
  \hline
  1.&$\{0_{A},1_{A},0_{B},1_{B}\}$& $\frac{1}{2}(1-c_6)$ & $0$ &$\frac{1}{2}(1-c_6) $ & $0$ & $0$ &$c_6$ \\
  \hline
  2.&$\{0_{A},1_{A},0_{B}\}$ & $c_1$ & $ \frac{1}{2}(1-c_6)-c_1$ & $\frac{1}{2}(1-c_6)$ & $0$ &$0$ & $c_6$ \\
  \hline

   3.&$\{0_{A},1_{A},1_{B}\}$ & $\frac{1}{2}(1-c_6)$ & $0$ & $\frac{1}{2}(1-c_6)$ & $0$ & $0$ & $c_6$ \\
   \hline
  4.&$\{0_{A},0_{B},1_{B}\}$ & $\frac{1}{2}(1-c_6)$ & $0$ & $c_3$ & $\frac{1}{2}(1-c_6)-c_3$ & $0$ & $c_6$ \\
  \hline
  5.&$\{1_{A},0_{B},1_{B}\}$& $\frac{1}{2}(1-c_6)$ & $0$ & $\frac{1}{2}(1-c_6)$ & $0$ & $0$ & $c_6$ \\
  \hline
  6.&$\{0_{A},1_{A}\}$ & $c_1$ & $\frac{1}{2}(1-c_6)-c_1$ & $\frac{1}{2}(1-c_6)$ & $0$ & $0$ & $c_6$ \\
  \hline
7.&$\{0_{B},1_{B}\}$ & $\frac{1}{2}(1-c_6)$ & $0$  & $c_3$  & $\frac{1}{2}(1-c_6)-c_3$  & $0$  & $c_6$  \\
  \hline
  8.&$\{1_{A},1_{B}\}$& $\frac{1}{2}(1-c_6)$ & $0$ & $\frac{1}{2}(1-c_6)$ & $0$ & $0$ & $c_6$ \\
  \hline
  9.&$\{0_{A},0_{B}\}$ & $c_1$ & $\frac{1}{2}(1-c_6)-c_1$ & $c_3$ & $\frac{1}{2}(1-c_6)-c_3$ & $0$ &$c_6$ \\
  \hline
  10.&$\{0_{A},1_{B}\}$ & $\frac{1}{2}(1-c_6)$ & $0$ & $c_3$ &$c_4$ & $\frac{1}{2}(1-c_6)-c_3-c_4$ & $c_6$ \\
  \hline
   11.&$\{1_{A},0_{B}\} $ & $c_1$ & $c_2$ & $\frac{1}{2}(1-c_6)$ & $0$ & $\frac{1}{2}(1-c_6)-c_1-c_2$ & $c_6$ \\
  \hline

 12.&$\{0_{A}\}$& $c_1$  & $\frac{1}{2}(1-c_6)-c_1$  & $c_3$  & $c_4$  & $\frac{1}{2}(1-c_6)-c_3-c_4$  & $c_6$  \\
  \hline
  13.&$\{1_{A}\}$ & $c_1$  & $\frac{1}{2}(1-c_6)-c_1-c_4-c_5$  & $\frac{1}{2}(1-c_6)$  & $c_4$  & $c_5$  & $c_6$  \\
  \hline
   14.&$\{0_{B}\}$ & $c_1$  & $\frac{1}{2}(1-c_6)-c_1-c_5$  & $c_3$  & $\frac{1}{2}(1-c_6)-c_3$  & $c_5$  & $c_6$  \\
  \hline
   15.&$\{1_{B}\}$& $\frac{1}{2}(1-c_6)$  & $\frac{1}{2}(1-c_6)-c_3-c_4-c_5$  & $c_3$  & $c_4$  & $c_5$  & $c_6$  \\
  \hline
\end{tabular}
\end{center}
\caption{For the no-signaling bipartite Hardy's correlation with
two dichotomic observable on either side, here each row gives the form
of solutions for the corresponding choice of inputs to be locally
random.}
\end{table}
}\end{widetext}

\subsection{Hardy's correlation respecting information causality}
Let us first briefly discuss the principle of information
causality (IC) \cite{IC}, then we would apply it in our study of
the property of local randomness for two input- two output Hardy's
nonsignaling correlations. IC principle states that for two
parties Alice and Bob, who are separated in space, the information
gain that Bob can reach about a previously unknown to him data set
of Alice, by using all his local resources and $m$ classical bit
communicated by Alice, is at most $m$ bits. This principle can be
well formulated in terms of a generic information processing task
in which Alice is provided with a $N$ random bits $\vec {a} =
(a_1, a_2,....., a_N)$ while Bob receives a random variable $b \in
\{1, 2,, ..., N\}$. Alice then sends $m$
 classical bits to Bob, who must output a single bit $\beta$ with the aim
 of guessing the value of Alice's b-th bit $a_b$. Their degree of
success at this
 task is measured by $$ I \equiv \sum_{K = 1}^{N} {I (a_K : \beta |b = K)},$$
 where $I (a_K : \beta |b = K)$ is Shannon mutual information between $a_K$ and $\beta$.
  Then the principle of information causality says that physically allowed theories must have
$I \le m$. The result that both classical and quantum correlations
satisfy this condition was proved in \cite{IC}. It was further
shown there that, if Alice and Bob share arbitrary two input-two
output nonsignaling correlations corresponding to conditional
probabilities $P(ab|xy)$, then by applying a protocol by van Dam
\cite{i} and Wolf and Wullschleger \cite{wolf}, one can derive a
necessary condition for respecting the IC principle. This
necessary condition reads,
\begin{equation}
\label{ic} E^{2}_1 + E^{2}_2 \leq1,
\end{equation}
where $E_j= 2P_j-1$ ($j=1,2$), and $P_1$, $P_2$ are defined by,
\begin{eqnarray}
\label{pipii}
P_1&=& \frac{1}{2}\left[p_{(a=b|00)} + p_{(a=b|10)}\right]\nonumber\\
   &=& \frac{1}{2}\left[p_{00|00} + p_{11|00} + p_{00|10} + p_{11|10}\right]\nonumber\\
P_2&=& \frac{1}{2}\left[p_{(a=b|01)} + p_{(a\neq b|11)}\right]\nonumber\\
   &=& \frac{1}{2}\left[p_{00|01} + p_{11|01} + p_{01|11} + p_{10|11}\right]
\end{eqnarray}
Here it is important to note that the condition (\ref{ic}) is only
a necessary condition (based on the protocol give in \cite{IC})
for respecting the IC principle. So a violation of
(\ref{ic})implies a violation of IC but the converse may not be
true. In fact, it is shown in \cite{allcock} that there are
examples where the condition (\ref{ic}) is satisfied but not the
IC. We now derive some one way implications about the property of
local randomness for two input - two output Hardy's nonsignaling
correlations. It is easy to verify that restricting Hardy's
nonsignaling correlations by condition (\ref{ic}) and
interchanging the roles of Alice and Bob we get,

\bea c^2_6+2(c_4+c_5)c_6+2(c_4+c_5)(c_4+c_5-1)\leq0\eea

\bea c^2_6+2(c_2+c_5)c_6+2(c_2+c_5)(c_2+c_5-1)\leq0\eea

By applying these conditions for all possible choices of inputs
that can be locally random for Hardy's nonsignaling correlations
(TABLE II), we get that at least one of the above two conditions
are violated for the cases $1-8$ but for the cases $9-15$ we can
find $c_{i}$s satisfying the above two conditions. Thus for the
cases $1-8$ we can conclude that IC is violated, hence they cannot
be true in quantum mechanics also. Now we shall study the cases
9-15 in the context of quantum mechanics in the following
subsection.

\subsection{Hardy's correlation in quantum mechanics}
Violation of IC for cases 1-8 implies that there are no quantum
solution for these cases. To resolve the remaining cases (9-15),
we consider a two qubit pure quantum state. It is to be
mentioned that for two qubits, Hardy's argument runs only for pure
entangled state \cite{gkar}. So without loss of any generality, we
consider the following two qubit state,
 \bea |\Psi
\rangle=\cos\beta
|0\rangle_{A}|0\rangle_{B}+\exp(i\gamma)\sin\beta|1\rangle_{A}|1\rangle_{B}\eea.
Then the density matrix $\rho_{AB}= |\Psi \rangle \langle \Psi|$
can be written in terms of Pauli matrices as,

\begin{widetext}
\bea \rho_{AB}=\frac{1}{4}[I^{A}\otimes
I^{B}+(cos^2\beta-\sin^2\beta)I^{A}\otimes\sigma^{B}_{z}+(cos^2\beta-\sin^2\beta)\sigma^{A}_{z}\otimes
I^{B}+
(2\cos\beta\sin\beta )\sigma^{A}_{x}\otimes\sigma^{B}_{x}\nonumber\\
+(2\cos\beta\sin\beta )\sigma^{A}_{x}\otimes\sigma^{B}_{y}+
(2\cos\beta\sin\beta
)\sigma^{A}_{y}\otimes\sigma^{B}_{x}-(2\cos\beta\sin\beta
)\sigma^{A}_{y}\otimes\sigma^{B}_{y}+
\sigma^{A}_{z}\otimes\sigma^{B}_{z}]\eea
\end{widetext}

The reduced density matrices $\rho_{A}$ and $\rho_{B}$ are,

\bea \rho_{A}= \frac {1}{2}[I
+(cos^2\beta-\sin^2\beta)\sigma^{A}_{z} ]\eea

\bea \rho_{B}= \frac {1}{2}[I
+(cos^2\beta-\sin^2\beta)\sigma^{B}_{z} ] \eea In general an
observable on a single qubit can be written as
$\hat{n}\cdot\sigma$ where, $\hat{n}= (\sin\theta \cos \phi
,\sin\theta \sin \phi, \cos \theta )$ is any unit vector in
$\mathbb{R}^{3}$ and $\sigma = (\sigma_{x},\sigma_{y},\sigma_{z})$. Then the projectors on the
eignestates of these observable are,

\bea P^{\pm}= \frac{1}{2}[I \pm\hat{n}\cdot\sigma]\eea

For observable on Alice's side to be locally random, \bea Tr
(\rho_{A}P^{+})= Tr (\rho_{A}P^{-})\eea similarly for observable
on Bob's side to be locally random, \bea Tr (\rho_{B}P^{+})= Tr
(\rho_{B}P^{-})\eea  On simplifying this we find that, for a
non-maximally entangled state an observable is locally random if
and only if $\theta = \frac{\pi}{2}$ i.e. $\hat{n}$ is of the form
$(\cos\phi , \sin\phi , 0)$. Here we would also like to mention
that for a maximally entangled state any arbitrary observable
shows the property of local randomness, but we know that Hardy's
argument doesn't run for a maximally entangled state. This also
follows from the IC principle, as for a maximally entangled state
any four arbitrary observable (two on Alice's side and two on
Bob's side) are locally random and we saw that if so, it violates
the IC principle.

 Now suppose $A$ ($0_{A}$) and
$A'$ ($1_{A}$) are the observable on Alice's side and $B$
($0_{B}$) and $B'$ ($1_{B'}$) are the observable on bob's side.
Here outputs $0$ and $1$ will corresponds to outcomes $+1$ and
$-1$ respectively. Then the Hardy's correlation can be written as,
\begin{widetext}
\bea
 P(A=+1,B=+1)=\cos^2\beta \cos^2\frac{\theta_{A}}{2}\cos^2\frac{\theta_{B}}{2}+\sin^2\beta \sin^2\frac{\theta_{A}}{2}\sin^2\frac{\theta_{B}}{2}\nonumber\\
+2\cos\beta\sin\beta
\sin\frac{\theta_{A}}{2}\sin\frac{\theta_{B}}{2}\cos\frac{\theta_{A}}{2}cos\frac{\theta_{B}}{2}\nonumber\\
\cos(\phi_{A}+\phi_{B}-\gamma)=0\eea
  \bea
 P(A=-1,B^\prime=-1)=\cos^2\beta \sin^2\frac{\theta_{A}}{2}\sin^2\frac{\theta_{B^\prime}}{2}+\sin^2\beta cos^2\frac{\theta_{A}}{2}cos^2\frac{\theta_{B^\prime}}{2}\nonumber\\
+2\cos\beta\sin\beta
\sin\frac{\theta_{A}}{2}\sin\frac{\theta_{B^\prime}}{2}\cos\frac{\theta_{A}}{2}\cos\frac{\theta_{B^\prime}}{2}\nonumber\\
\cos(\phi_{A}+\phi_{B^\prime}-\gamma)=0\eea \bea
 P(A^\prime=-1,B=-1)=\cos^2\beta \sin^2\frac{\theta_{A^\prime}}{2}\sin^2\frac{\theta_{B}}{2}+\sin^2\beta \ \cos^2\frac{\theta_{A^\prime}}{2}\cos^2\frac{\theta_{B}}{2}\nonumber\\
+2\cos\beta\sin\beta
\sin\frac{\theta_{A^\prime}}{2}\sin\frac{\theta_{B}}{2}\cos\frac{\theta_{A^\prime}}{2}\cos\frac{\theta_{B}}{2}\nonumber\\
\cos(\phi_{A^\prime}+\phi_{B}-\gamma)=0\eea

\bea
P(A^\prime=-1,B^\prime=-1)=\cos^2\beta \sin^2\frac{\theta_{A^\prime}}{2}\sin^2\frac{\theta_{B^\prime}}{2}+\sin^2\beta \cos^2\frac{\theta_{A^\prime}}{2}\cos^2\frac{\theta_{B^\prime}}{2}\nonumber\\
+2\cos\beta\sin\beta
sin\frac{\theta_{A^\prime}}{2}\sin\frac{\theta_{B^\prime}}{2}cos\frac{\theta_{A^\prime}}{2}\cos\frac{\theta_{B^\prime}}{2}\nonumber\\
\cos(\phi_{A^\prime}+\phi_{B^\prime}-\gamma)\neq0\eea
\end{widetext}
For these Hardy's correlation if observable $A$ and $B$ ($0_{A}$
and $0_{B}$) are locally random, then
$\theta_{A}=\theta_{B}=\frac{\pi}{2}$, then from equation $(15)$
we get, \bea 1 + \sin 2\beta \cos(\phi_{A}+\phi_{B}-\gamma)=0\eea
then this equation is satisfied only if $\sin 2\beta$ takes the value $+1$ or $-1$, in either case
corresponding state has to be a maximally entangled state, but this cannot be a case.
Therefore we conclude that observable $A$ and $B$ cannot be locally
random in quantum mechanics. Similarly we can see that local
randomness of two observable in the cases, $A'$ and $B$ ($1_{A}$
and $0_{B}$) and $A$ and $B'$ ($0_{A}$ and $1_{B}$) is also not
possible.\\

Now we consider the case of just one observable - say $A$$(0_{A})$
from the set $\{A,A',B,B'\}$ to be locally random ( and similarly
for the cases $A',B,B'$). Then we find that there are
non-maximally entangled states and choices of observable
$A,A',B,B'$ such that one of the observable is locally
random. We give an example, consider the state $\beta =
\frac{\pi}{6}$, and $\gamma=\pi$, choose observable $A$ as
$\theta_{A}=\frac{\pi}{2}$ and $\phi_{A}=\pi$, $A'$ as
$\theta_{A'}=2\tan^{-1}(\tan^{2}\frac{\pi}{6})$ and
$\phi_{A'}=-\pi$, $B$ as $\theta_{B}=\frac{2\pi}{3}$ and
$\phi_{B}=\pi$, and $B'$ as $\theta_{B'}=\frac{\pi}{3}$ and
$\phi_{B'}=-\pi$, then it can be easily checked that for this
choice of state and observable, Hardy's argument runs and the
observable $A$ is locally random. Thus by analyzing the remaining
cases $(9-15)$ within quantum mechanics, we can now conclude that
for a quantum mechanical state showing Hardy's nonlocality, at most
one out of the four observable can be locally random.

\section{Conclusion}
Maximally entangled state in quantum mechanics does not reproduce Hardy's correlation
whereas generalized non-signaling theory put no such restriction on
the local randomness of the observable for Hardy's correlation. We study all the
possibilities of local randomness in Hardy's correlation in the context of information causality condition.
 We observe that not only in terms of value of maximal probability of success \cite{ashu}, but also in term of local randomness
 there is gap between quantum mechanics and information causality condition. It remains to see, in future,  whether
 some stronger necessary condition for information causality can close this gap.

\begin{acknowledgments}
It is a pleasure to thank Guruprasad Kar and Sibasish Ghosh for
many stimulating discussions. AR acknowledge support from DST
project SR/S2/PU-16/2007. RR acknowledge support from Norwegian
Research Council.
\end{acknowledgments}


\begin{thebibliography}{99}

\bibitem{bell} J.S. Bell, Physics {\bf 1}, 195 (1964); J.F. Clauser, M.A. Horne, A. Shimony, and R.A.
Holt,  Phys. Rev. Lett. {\bf 23}, 880 (1969).

\bibitem{ghirardi} G.C. Ghirardi, A. Rimini and T. Weber,  Lett. Nuovo
Cim. {\bf 27} (1980) 263.

\bibitem{PR} S. Popescu and D. Rohrlich, Found. Phys. {\bf
24}, 379 (1994); Sandu Popescu, arXiv:quant-ph/9709026 (1997).



\bibitem{a} L. Masanes, A. Acin, and N. Gisin, Phys. Rev. A {\bf 73}, 012112 (2006).
\bibitem{b} H. Barnum, J. Barret, M. Leifer, and A. Wilce, Phys. Rev. Lett.{\bf 99}, 240501 (2007).
\bibitem{c} V. Scarani {et al.} Phys. Rev. A {\bf 74}, 042339 (2006).
\bibitem{d} J. Barret, L. Hardy, and A. Kent,  Phys. Rev. Lett. {\bf 95}, 010503 (2005).
\bibitem{e} A. Acin, N. Gisin, and L. Masanes, Phys. Rev. Lett. {\bf 97}, 120405 (2006).
\bibitem{f} L. Masanes, Phys. Rev. Lett. {\bf 102}, 140501 (2009).
\bibitem{g} H. Barnum, J. Barret, M. Leifer, and A. Wilce, arXiv:quant-ph/0805.3553v1 (2008).
\bibitem{h} P. Skrzypczyk, N. Brunner, and S. Popescu, Phys. Rev. Lett. {\bf 102}, 110402 (2009).
\bibitem{i} W. van Dam, e-print  arXiv:quant-ph/0501159.
\bibitem{j} G. Brassard,  Phys. Rev. Lett. {\bf 96}, 250401 (2006).
\bibitem{k} N. Linden, S. Popescu, A.J. Short, A. Winter, Phys. Rev. Lett. {\bf 99}, 180502 (2007).
\bibitem{l} N. Brunner, and P. Skrzypczyk, Phys. Rev. Lett. {\bf 102}, 160403 (2009).
\bibitem{m} J. Barrett, Phys. Rev. A {\bf 75}, 032304 (2007).
\bibitem{n} J. Barrett, N. Linden, S. Massar, S. Pironio, S. Popescu and D. Roberts  Phys. Rev. A {\bf 71},
022101 (2005).

\bibitem{IC} M. Pawlowski,T. Paterek, D. Kaszlikowski, V. Scsrani,
A. Winter and M. Zukowski,  Nature {\bf 461}, 1101 (2009).

\bibitem{tsirelson} B.S. Tsirelson, Lett. math. Phys. {\bf
4}, 93 (1980).

\bibitem{allcock} Jonathan Allcock, Nicolas Brunner, Marcin
Pawlowaski, and Valerio Scarani, Phys. Rev. A {\bf 80},
040103(R)(2009).
\bibitem{hardy92} L. Hardy, Phys. Rev. Lett. {\bf 68}, 2981 (1992).
\bibitem{hardy93} L. Hardy, Phys. Rev. Lett. {\bf 71}, 1665 (1993).
\bibitem{hardyns} S.K. Chaudhary, S. Ghosh, G. Kar, S. Kunkri, R. Rahaman, and A. Roy, e-print  arXiv:0807.4414.
\bibitem{hardyqm} S. Kunkri, S.K. Chaudhary, A. Ahanj, and P. Joag, Phys. Rev. A {\bf 73}, 022346 (2006).

\bibitem{ashu} Ali Ahanj, Samir Kunkri, Ashutosh Rai, Ramij Rahaman, and Pramod S. Joag,  Phys. Rev. A {\bf 81}, 032103 (2010).
\bibitem{cabello} Adan Cabello,  Phys. Rev. A {\bf 61}, 022119 (2000).
\bibitem{wolf} S. Wolf and J. Wullschleger, e-print  arXiv:quant-ph/0502030v1 (2005).
\bibitem{gkar} G. Kar, Phys. Lett. A {\bf 228}, 119 (1997).
\end{thebibliography}
\end{document}